# Dynamic Stiction Mode by Friction Vector Rotation

Ken Nakano[1] & Valentin L. Popov[2]
[1]Yokohama National University, Yokohama 240-8501, Japan
[2]Technische Universität Berlin, Berlin 10623, Germany

We numerically study a simple sliding system: a rigid mass pulled by a spring with a strong in-plane stiffness anisotropy and a small misalignment angle. Simulations show that the apparent stick phase appearing in this system is in reality a phase of very slow creep, followed by a rapid sliding, slip. Surprisingly, the absolute value of the friction force remains almost constant from the very beginning of the stick phase, merely rotating in the sliding plane. We call this specific mechanism of apparent stick due to rotation of the force vector "dynamic stiction".

*Introduction*.–Dry friction plays an essential role in many physical processes and numerous engineering, biological, and medical applications [1]-[3]. Often, the simplest friction "law" (Amontons' law [4],[5]) is used to describe dry friction. It states that an object remains at rest until a driving force exceeds some critical value called static friction. Thereafter, sliding occurs at an almost constant force $F$, which is roughly proportional to the normal force $W$: $F = \mu W$. The proportionality coefficient $\mu$ is called the friction coefficient. At the latest since Coulomb [6], one distinguishes between *static* and *kinetic* friction. This study is devoted to the analysis of the *transition* between *static* and *kinetic* friction or, in other words, *transition* from *stick* to *slip*. This transition plays an essential role in many technological [7] and geological [8] processes and has been studied intensively along two approaches, namely, *static* and *kinetic*. In the former approach, the tangentially loaded contact is considered as being divided into sliding and sticking parts, while at the increasing tangential load, the fraction of stick decreases until complete sliding begins. The prominent representative of this approach is the theory of partial sliding by Cattaneo [9] and Mindlin [10]. In the latter approach, both stick and slip are considered as sliding with different velocities. The prominent representatives of this approach are the rate- and state-dependent laws of friction developed in the 1970s in the context of geotectonic [11],[12]. A similar but a purely phenomenological approach has been developed in the context of *pre-sliding* [13],[14] which is of basic importance for precision positioning and feedback control systems [15]. Further approaches combining both perspectives have been developed, e.g., rapid propagation of detachment fronts in a contact plane [16].

All of the above approaches are entirely focused on describing the *magnitude* of the friction force. Its *direction* is assumed to be opposite to that of apparent sliding. However, this is not necessarily true owing to the finite stiffness of any real sliding system. Considering the *direction* of the friction force introduces an additional degree of freedom and opens a completely new view on the old problem of transition from stick to slip. Here we show that both stick and slip phases can be naturally understood in a purely mechanical way as emerging from the *rotation* of the friction vector. The importance of the rotation of the friction vector was highlighted in [17] in context of active control of friction by transverse oscillations.

*In-plane misalignment*.–The most crucial concept in this study is the inevitable existence of *in-plane misalignment*. For example, many researchers have used pin-on-plate-type apparatus in laboratory friction tests (Fig. 1). A pin mounted to a cantilever spring is in contact with a plate driven at a velocity $V$. To induce a measurable deflection of the spring, the stiffness in the drive direction, $k_x$, is designed to be sufficiently low, whereas that in the perpendicular direction, $k_y$, is usually several orders of magnitude higher. However, due to fabrication or setting errors and finite deformations in use, perfect elimination of the misalignment between the principal axes of the stiffness tensor and the drive direction is practically impossible. A careful setting could make the misalignment angle $\varphi$ considerably small (e.g., less than 1°) and can be neglected in measuring the magnitude of friction in steady states. However, as will be shown in this Letter, the influence even of such small misalignment is *not negligible* in the non-steady dynamics and plays a vital role in inducing friction vector rotation.

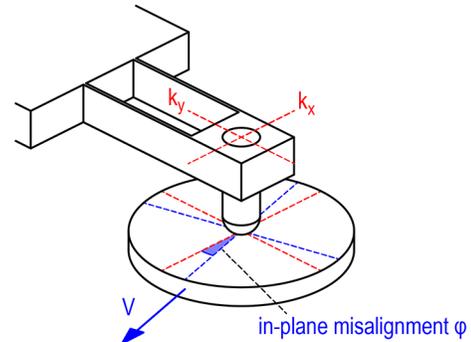

FIG. 1. Typical structure of apparatus for friction tests. In real systems, the in-plane misalignment $\varphi$ is inevitable between the principal axes of the spring stiffness (red) and the drive direction (blue).

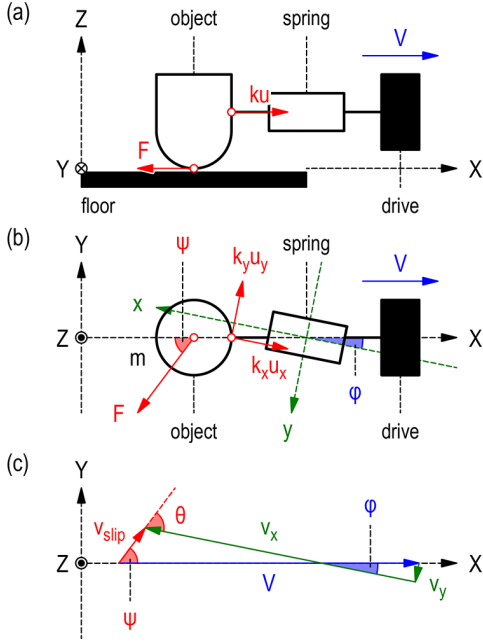

FIG. 2. Analytical model. (a) Side view: An object in contact with a stationary floor is pulled to the right via a spring. (b) Top view: An in-plane misalignment $\varphi > 0$ exists between the pulling direction $X$ and the principal axis $x$ of the stiffness tensor. (c) Geometrical relationship of velocities (drive velocity $V$, spring elongation rates $v_x$ ($=\dot{u}_x$) and $v_y$ ($=\dot{u}_y$), and slip velocity $v_{\text{slip}}$) and angles ($\varphi$, $\psi$, and $\theta$). The angle $\psi$ of the friction force $\mathbf{F}$ in (b) is determined from the direction of $\mathbf{v}_{\text{slip}}$ in (c). Note that $\theta = \psi + \varphi$.

*Methods.*–Consider a mass-spring system with constant in-plane misalignment (Fig. 2). In the side view (Fig. 2(a)), an object is coupled to a spring and in contact with a stationary floor ($XY$ plane). The right end of the spring is driven along the $X$-axis with a velocity $V$. In the top view (Fig. 2(b)), the spring is characterized by a stiffness tensor. Let $x$ and $y$ be the principal axes of the stiffness tensor, and $k_x$ and $k_y$ be the corresponding principal values. We assume a small misalignment angle $\varphi > 0$ between the $x$- and $X$-axes. The projections of the spring force onto the principal axes are $k_x u_x$ and $k_y u_y$, where $u_x$ and $u_y$ are the spring elongations in the $x$ and $y$ directions, respectively. Letting $\mathbf{v}_{\text{slip}}$ be the slip velocity vector of the object, the direction of the friction vector $\mathbf{F}$ was assumed to be opposite to that of $\mathbf{v}_{\text{slip}}$ and its magnitude a function of the magnitude of $\mathbf{v}_{\text{slip}}$: $F = F(v_{\text{slip}})$. A velocity-weakening law $F = [\mu_\infty + (\mu_0 - \mu_\infty) \exp(-v_{\text{slip}}/v_f)]W$ was used, where $\mu_0$ and $\mu_\infty$ are the friction coefficients for $v_{\text{slip}} \approx 0$ and $\infty$, respectively, and $v_f$ and $W$ are the velocity constant and the normal load, respectively. Note that we *do not* assume the existence of a finite *static* friction (which vanishes).

The equations of motion of the object with a mass $m$ in the coordinates $x$ and $y$ are as follows:

$$m\ddot{u}_x + k_x u_x = F\cos\theta, \quad m\ddot{u}_y + k_y u_y = F\sin\theta, \quad (1)$$

where $\theta = \psi + \varphi$ is the angle between $\mathbf{F}$ and the $x$-axis. From the geometry represented in Fig. 2(c), we obtain

$$\cos\theta = \frac{V\cos\varphi - \dot{u}_x}{v_{\text{slip}}}, \quad \sin\theta = \frac{V\sin\varphi - \dot{u}_y}{v_{\text{slip}}} \quad (2)$$

with

$$v_{\text{slip}} = \sqrt{(V\cos\varphi - \dot{u}_x)^2 + (V\sin\varphi - \dot{u}_y)^2}. \quad (3)$$

Using Eqs. (2) and (3), Eq. (1) can be rewritten as

$$m\ddot{u}_x + k_x u_x = F\frac{V\cos\varphi - \dot{u}_x}{\sqrt{(V\cos\varphi - \dot{u}_x)^2 + (V\sin\varphi - \dot{u}_y)^2}}, \quad (4)$$

$$m\ddot{u}_y + k_y u_y = F\frac{V\sin\varphi - \dot{u}_y}{\sqrt{(V\cos\varphi - \dot{u}_x)^2 + (V\sin\varphi - \dot{u}_y)^2}}. \quad (5)$$

These two non-linear second-order differential equations completely determine the dynamics of the system. They were solved numerically using the Runge-Kutta method. The time evolution of the object position in the laboratory coordinates can be obtained by

$$X = Vt - u_x \cos\varphi - u_y \sin\varphi. \quad (6)$$

*Results.*–Figure 3 presents solutions to the equations of motion. A small in-plane misalignment of $\varphi = 1°$ and a strong stiffness anisotropy of $k_y/k_x = 10^4$ were assumed with considering typical cantilever springs used in laboratory friction tests (Fig. 1). Other parameters are listed in the caption. The drive started to move at $t = 0$. The time evolutions of the $x$- and $y$-components of the spring force are shown in the top rows. The longitudinal component $k_x u_x$ shows the classical stick-slip behavior consisting of a linear increase in time followed by a rapid drop (Fig. 3(a)). The transverse component $k_y u_y$, in contrast, reveals a counterintuitive behavior. It jumps to the maximum value (equal to the magnitude of the friction force in slow sliding) and subsequently decreases to vanish at the beginning of the slip phase (Fig. 3(b)). The magnitude of the friction force remains practically constant during the whole stick phase (Fig. 3(c)), reducing only in the phases of rapid slip. Maintaining equilibrium in the pulling direction is possible due to the in-plane rotation of the friction vector described by the angle $\psi$ between the directions of the force and the drive direction (Fig. 3(d)). The time evolution of the longitudinal coordinate $X$ (Fig. 3(e)) shows a pronounced stick-slip character. Although the stair-like object position ($X$) and the sawtooth-shaped spring force ($k_x u_x$) indicate typical stick-slip, in reality, the object never comes to a full stop. During the stick phases, the object is slowly slipping and gradually accelerated in the $X$ direction (see the inset of Fig. 3(e)), which reminds us of the so-called "slow creep" known from studies on the rate- and state-dependent friction laws [18]. To underline the dynamic nature of the apparent stick phase we call it "dynamic stiction".

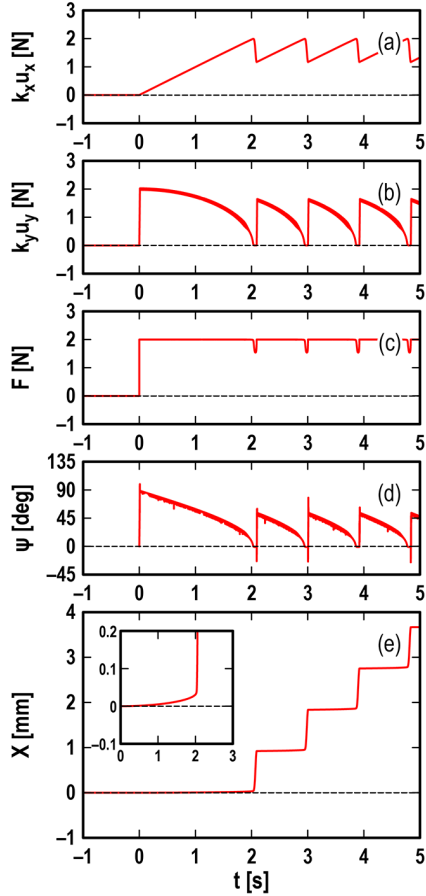

FIG. 3. Dynamic stiction, slow creep, and stick-slip. (a) Spring force, $k_x u_x$. (b) Spring force, $k_y u_y$. (c) Magnitude of friction force, $F$. (d) Direction of friction force, $\psi$. (e) Object position, $X$. System parameters: $\varphi = 1°$, $m = 0.25$ kg, $k_x = 1$ kN/m, $k_y = 10$ MN/m, $\mu_0 = 0.20$, $\mu_\infty = 0.15$, $v_f = 10$ mm/s, $W = 10$ N, and $V = 0$ for $t < 0$ and 1 mm/s for $t \geq 0$. Initial conditions: $X(0) = 0$, $Y(0) = 0$, $\dot{X}(0) = \varepsilon V$, and $\dot{Y}(0) = 0$, where $\varepsilon = 10^{-6}$.

Another interesting feature observed in the numerical simulations is the high-frequency oscillations (Fig. 4). These oscillations are not numerical artifacts but rather inherent properties of the system. Frequency analysis reveals that the oscillation frequency coincides at the beginning of the "stick" phase with the natural frequency of the transverse oscillations, $\omega_y = (k_y/m)^{1/2}$ (for the system parameters used, 1.0 kHz) and decreases when approaching the phase of rapid slip (Fig. 4(b) and (d)). The high-frequency dynamics of sliding systems is of significant interest for many technical applications [19], [20]. Although their physical origin and influencing factors have been studied for decades, their nature often remains unclear [19]. Thus, the concept of friction vector rotation presents a new perspective also on this problem.

In the following, we discuss in more detail the main features of the observed stick-slip motion: (a) *dynamic stiction*, (b) *slow creep*, (c) *high-frequency dynamics*, and (d) *low-frequency dynamics*.

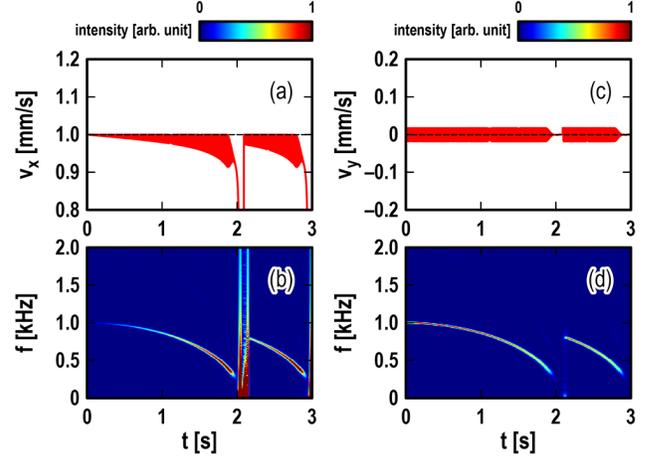

FIG. 4. High-frequency oscillations appearing in the "stick" phase. (a) Spring elongation rate, $v_x (= \dot{u}_x)$, and (b) its spectrogram. (c) Spring elongation rate, $v_y (= \dot{u}_y)$, and (d) its spectrogram. System parameters and initial conditions are the same as in Fig. 3. The spectrograms were obtained using short-time Fourier transform with Hamming window of the width 0.1 s.

*Dynamic stiction.*–Based on the results presented so far, we describe schematically the mechanism of dynamic stiction. The simulations revealed that the object, when loaded, never stops but is moving with a small velocity. This leads to the fact that the magnitude of the friction force remains constant at all time. At the beginning of sliding, the projection of the friction force on the drive direction gradually increases from zero. This means that the friction vector should be initially directed *perpendicular* to the drive direction. In fact, we observe that at the beginning of macroscopic sliding, the angle $\psi$ jumps to almost 90° (Fig. 3(d)). When the spring force in the drive direction increases, the friction vector rotates, but its absolute value $F = (F_x^2 + F_y^2)^{1/2}$ remains constant and is practically equal to the value of the kinetic friction at the nearly zero velocity, $F_0 = \mu_0 W$ (Fig. 3(c)). Therefore, the perpendicular component of the friction force $F_y = (F_0^2 - F_x^2)^{1/2}$ is decreasing, gradually approaching zero (Fig. 3(b)). The change in the pulling force can be supported by the friction vector rotation only until $F_x \leq F_0$. As soon as the pulling force exceeds this critical value, no static equilibrium is possible anymore, and the phase of rapid slip begins.

*Slow creep.*–Let us consider in more detail the "stick" phase (that in reality is the phase of slow creep). The movement in this stage is quasistatic, which means that the inertial terms are negligible. However, this is valid only for movement in the $x$-direction. The high transverse stiffness $k_y$ guarantees very small deflections $u_y$. Although the velocity $\dot{u}_y$ is not necessarily small due to high natural frequency in the $y$-direction, it has zero average and can be set to zero while considering the creep process. Therefore, in the creep phase, we can neglect the terms with $\ddot{u}_x$ and $\ddot{u}_y$ in Eqs. (4) and (5). After some transformations, this leads to

$$\dot{u}_x = V\cos\varphi - \frac{k_x u_x / F_0}{\sqrt{1-(k_x u_x / F_0)^2}} V\sin\varphi. \quad (7)$$

This is an ordinary differential equation of first order that completely determines the dynamics of the degree of freedom $u_x(t)$. The coordinate $X$ in the drive direction can finally be found using Eq. (6). The resulting solution shows that at small misalignment, the system shows an almost perfect stick, while it rapidly becomes blurred when increasing the misalignment angle. In the limiting case of very small misalignment angles, Eq. (7) takes the form $\dot{u}_x \approx V$ with the solution $u_x = Vt$ (for $Vt < l_x = F_0/k_x$). For the creep velocity, Eq. (6) yields

$$\dot{X} \approx V - \dot{u}_x \approx V\varphi\left(\varphi + \frac{(Vt/l_x)^2}{\sqrt{1-(Vt/l_x)^2}}\right). \quad (8)$$

It contains two contributions: One is linear in $V$ and of second order in $\varphi$, while the other is linear in $\varphi$ and of second order in $V$. The latter has a singularity of the form $(1-(Vt/l_x)^2)^{-1/2}$ when approaching the moment of "slip". Comparison with published measurements of creep [21] seems to confirm the existence of these two contributions.

*High-frequency dynamics.*–At the beginning of the creep phase, $V\cos\varphi - \dot{u}_x \approx 0$. Neglecting this term in Eq. (5) transforms this equation to

$$m\ddot{u}_y + k_y u_y + F_0 \text{sign}(\dot{u}_y - V\sin\varphi) = 0. \quad (9)$$

The average value of displacement is easily found by setting $\dot{u}_y = 0$ and $\ddot{u}_y = 0$: $k_y\langle u_y\rangle = F_0$. The amplitude of oscillations is determined by the non-linear term $F_0\text{sign}(\dot{u}_y - V\sin\varphi)$, which is bounded by the value $\bar{\dot{u}}_y = V\sin\varphi$. After a setting time, the system oscillates at around the high natural frequency $\omega_y = (k_y/m)^{1/2}$ that is observed as high-frequency oscillations (Fig. 4).

*Low-frequency dynamics.*–The low-frequency dynamics appearing in the x-direction can be described with a single equation by setting $\dot{u}_y = 0$ in Eq. (4). By selecting $\tau = \omega_x t$ as the new independent variable and $v_x = \dot{u}_x$ as the new dependent variable, under the simplification of $F = F_0$, the equation is reduced to the following simple form:

$$v_x'' + 2(A\sin^3\theta)v_x' + v_x = 0. \quad (10)$$

Note that $A = \mu_0 W/2V\sin\varphi(mk_x)^{1/2}$ is the single dimensionless parameter determined by six system parameters. Therefore, the stability of the low-frequency dynamics is controlled solely by the dimensionless damping factor $\zeta = A\sin^3\theta$. Recall that during the "stick" phase, the direction of the friction force gradually changes from $\theta = 90°$ toward $\varphi$ (Fig. 3(d)) represented in the damping diagram (Fig. 5) by the red and blue vertical lines for $\varphi = 1°$ and $10°$, respectively. The small misalignment ($\varphi = 1°$) initially provides an overdamped state with a high damping factor ($\zeta = A = 3.6\times10^3$ when $\theta = 90°$), leading to a slow slip with low acceleration (i.e., the slow creep). Thereafter, by crossing the critical boundary represented by the green line for $\zeta = 1$, the system finally turns into an underdamping state ($\zeta = 1.9\times10^{-2}$ when $\theta = \varphi = 1°$), leading to a fast slip as a result of the free oscillation at around the low natural frequency $\omega_x = (k_x/m)^{1/2}$. Note that the larger misalignment ($\varphi = 10°$) prevents the transition from the overdamping to underdamping states, which could be recognized as the stabilization effect of the in-plane misalignment to suppress friction-induced oscillations [22],[23]. Also note that a system with $A < 1$ (e.g., a system with large $\varphi$, large $m$, large $k_x$, small $\mu_0$, small $W$, and large $V$) takes the underdamping state from the beginning, in which the dynamic stiction would disappear, which will be discussed elsewhere.

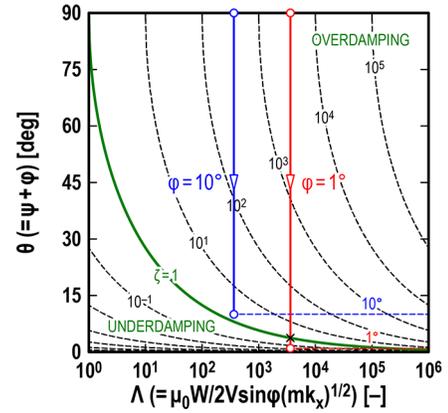

FIG. 5. Damping diagram for the low-frequency dynamics: The instantaneous damping ratio $\zeta = A\sin^3\theta$ is mapped in the plot of $\theta$ versus $A$. A small misalignment $\varphi = 1°$ causes the transition from the overdamping- to underdamping-states at the cross mark, which is observed as the transition from slow- to fast-slip (i.e., the stick-to-slip transition). Note that a larger misalignment $\varphi = 10°$ suppresses the transition. System parameters: $m = 0.25$ kg, $k_x = 1$ kN/m, $\mu_0 = 0.20$, $W = 10$ N, and $V = 1$ mm/s.

*Conclusion.*–Our results establish a new perspective of slow creep, stick-to-slip transition, and the nature of high-frequency oscillations in sliding systems. We are completely aware that the demonstrated mechanism of stick-to-slip transition does not exhaust all possible mechanisms of the stick-slip phenomenon. However, we would like to draw attention of researchers and engineers to the fact that the well-known and highly debated properties of the transition from stick to slip, including slow creep, may have a completely different – and much simpler – purely mechanical origin.

[1] F. P. Bowden and D. Tabor, *The Friction and Lubrication of Solids* (Clarendon Press, 1986).
[2] B. N. J. Persson, *Sliding Friction. Physical Principles and Applications* (Springer, 2000).
[3] V. L. Popov, *Contact Mechanics and Friction. Physical Principles and Applications*. 2nd Ed. (Springer, 2017).
[4] G. Amontons, Mem. l'Academie Royale, 206 (1699).
[5] E. Popova and V. L. Popov, *Friction* **3**, 183 (2015).

[6] C. A. Coulomb, *Theorie des machines simple*. (Bachelier, 1821).
[7] B. Feeny, A. Guran, N. Hinrichs, and K. Popp, *Appl. Mech. Rev.* **51**, 321 (1998).
[8] C. H. Scholz, *Nature* **391**, 37 (1998).
[9] C. Cattaneo, *Rendiconti Dell'Acad. Nazionale Dei Lincei* **27**, 342, 434, 474 (1938).
[10] R. D. Mindlin, *J. Appl. Mech.* **16**, 259 (1949).
[11] J. H. Dieterich, *Pure Appl. Geophys.* **116**, 790 (1978).
[12] J. R. Rice and A. L. Ruina, *J. Appl. Mech.* **50,** 343–349 (1983).
[13] C. C. de Wit, H. Olsson, K. Astrom, and P. Lischinsky, *IEEE Trans. Autom. Contr.* **40**, 419 (1995).
[14] V. Lampaert, F. Al-Bender, and J. Swevers, *Tribol. Lett.* **16,** 95 (2004).
[15] B. Armstrong-Hélouvry, P. Dupont, and C. C. de Wit, *Automatica* **30**, 1083 (1994).
[16] S. M. Rubinstein, G. Cohen, and J. Fineberg, *Nature* **430**, 1005 (2004).
[17] J. Benad, K. Nakano, V. L. Popov, and M. Popov, *Friction* **7**, 74 (2018).
[18] V. L. Popov, B. Grzemba, J. Starcevic, and M. Popov, *Tectonophys.* **532**, 291 (2012).
[19] N. M. Kinkaid, O. M. O'Reilly, and P. Papadopoulos, *J. Sound Vib.* **267**, 105 (2003).
[20] J. T. Oden and J. A. C. Martins, *Comput. Methods Appl. Mech. Eng.* **52**, 527 (1985).
[21] B. Grzemba, *Predictability of Elementary Models for Earthquake Dynamics*. (epubli GmbH, Berlin, 2014).
[22] N. Kado, C. Tadokoro, and K. Nakano, *Trans. Jpn. Soc. Mech. Eng. Ser. C* **79**, 396 (2013).
[23] N. Kado, C. Tadokoro, and K. Nakano, *Tribol. Online* **9**, 63 (2014).